\documentclass[a4paper,12pt]{article}
\usepackage{graphicx}
\usepackage{amsmath,amsthm,amssymb,mathrsfs,mathtools}
\usepackage{xr-hyper}

\usepackage{anyfontsize}
\usepackage[OT1]{fontenc} 
\usepackage{etoc}
\usepackage{float}
\usepackage{scalerel}
\usepackage{fancyhdr}

\def\subsectiontitle{}
\def\subsubsectiontitle{}
\usepackage[font=small,margin=0pt,skip=0pt]{caption}

\usepackage{tikz}
\usetikzlibrary{arrows,automata, positioning}
\usetikzlibrary{arrows.meta}
\tikzset{
  >={To[length=2.5pt]}
  }
\usepackage{mleftright}
\mleftright
\usepackage{mathpazo,mathabx}
\usepackage{enumitem}

\usepackage[margin=1in]{geometry}
\usepackage[pdfencoding=auto,psdextra,colorlinks,breaklinks=true]{hyperref}
\usepackage{bookmark}
\usepackage{breakcites}

\usepackage{xcolor}
\AtBeginDocument{%
	\hypersetup{%
    linkcolor={red!50!black},%
    citecolor={blue!50!black},%
		filecolor={blue!80!white},
    urlcolor={blue!80!black}%
}%
}%

\usepackage[nameinlink,noabbrev,sort,capitalise]{cleveref}

\usepackage{chngcntr}
\usepackage{apptools}
\AtAppendix{\counterwithin{lem}{section}}

\newtheorem{prop}{Proposition}
\crefname{prop}{Proposition}{Propositions}

\newtheorem{thm}{Theorem}
\crefname{thm}{Theorem}{Theorems}

\crefname{cor}{Corollary}{Corollaries}

\newtheorem{lem}{Lemma}
\crefname{lem}{Lemma}{Lemmas}

\newtheorem{ass}{Assumption}
\crefname{ass}{Assumption}{Assumptions}

\crefname{assalp}{Assumption}{Assumptions}

\newtheorem{defi}{Definition}
\crefname{defi}{Definition}{Definitions}

\theoremstyle{remark}
\newtheorem*{remark}{Remark}

\theoremstyle{definition}
\newtheorem{eg}{Example}
\crefname{eg}{Example}{Examples}

\crefname{problem}{Problem}{Problems}

\newcommand{\overbar}[1]{\mkern 1.5mu\overline{\mkern-1.5mu#1\mkern-1.5mu}\mkern 1.5mu}

\usepackage{indentfirst}

\allowdisplaybreaks

\let\oldfootnote\footnote
\renewcommand\footnote[1]{\oldfootnote{\hspace{.4mm}#1}}



\let\oldFootnote\footnote
\newcommand\nextToken\relax

\renewcommand\footnote[1]{%
    \oldFootnote{#1}\futurelet\nextToken\isFootnote}

\newcommand\isFootnote{%
    \ifx\footnote\nextToken\textsuperscript{,}\fi}

\def\d{\mathrm{d}}

\usepackage[it,small,compact]{titlesec}
\titleformat*{\section}{\bfseries}
\titlespacing*{\section}{0.5em}{0.5em}{0pt}
\titlespacing*{\subsection}{0.3em}{0.3em}{0pt}
\titlespacing*{\subsubsection}{0.2em}{0.2em}{0pt}

\usepackage{titling}
\pretitle{\begin{center}\vspace{-3em}\Large}
\preauthor{\begin{center}\vspace{-1.5em}\normalsize\begin{tabular}[t]{c}}
\predate{\begin{center}\vspace{-1em}\normalsize}

\providecommand{\keywords}[1]{\vspace{0.5em}\\ \noindent\textit{Keywords:} #1\\}
\providecommand{\jel}[1]{\noindent \textit{JEL classification:} #1}

\renewenvironment{abstract}{\par\noindent\hrulefill\\ \textbf{\abstractname.}\ \ignorespaces}{\vskip -0.5em \noindent\hrulefill\par\medskip}

\usepackage{natbib}

\title{Selling Information \thanks{ I thank Alessandro Bonatti, Yeon-Koo Che, Navin Kartik, Qingmin Liu, Konrad Mierendorff, Mike Riordan for helpful comments and suggestions. I also want to thank seminar participants at Columbia University Micro Theory Colloquium and CRETA at University of Warwick. All errors are mine.}}
\author{Weijie Zhong\thanks{\emph{Department of Economics, Columbia University}, Email: \url{wz2269@columbia.edu}}}
\date{Match, 2016}

\begin{document}
\maketitle

\begin{abstract}
I consider the monopolistic pricing of informational good. A buyer's willingness to pay for information is from inferring the unknown payoffs of actions in decision making. A monopolistic seller and the buyer each observes a private signal about the payoffs. The seller's signal is binary and she can commit to sell any statistical experiment of her signal to the buyer. Assuming that buyer’s decision problem involves rich actions, I characterize the profit maximizing menu. It contains a continuum of experiments, each containing different amount of information. I also find a complementarity between buyer’s private information and information provision: when buyer’s private signal is more informative, the optimal menu contains more informative experiments.
	\keywords{information design, mechanism design, monopoly pricing}
	\jel{D42, D82, D83}
\end{abstract}

\section{Introduction}
When an individual is making decision involving unknown payoffs, he often has the opportunity to acquire information to learn the payoffs. For example, an investor is choosing from a set of assets, whose returns are unknown. The investor can learn about the assets’ returns from multiple information sources: his own knowledge about the market, advices from financial consultants, articles on The Wall Street Journal, etc. Interestingly, the external information providers tend to offer increasingly richer menus of personalized options of information. For example, The Wall Street Journal and Bloomberg Businessweek are providing pay-by-article services through online platforms; the Amazon Web Services provides complicated pricing schedules for its cloud services. Meanwhile, although the menu is rich, each options provided in the menu are usually simple --- in previous examples of journal articles and data services they are often simple partial revelations of the complete information available from the information provider. \par

In this paper, those rich but simple menus of informational goods are justified as the revenue maximizing menus for the information seller, when the buyer has other private information sources unknown to the seller. I model the buyer (he) of information as a decision maker who is choosing from actions with unknown payoffs. The buyer and the seller (she) each observes a private signal about the payoffs. The seller can costlessly produce any product that contains weakly less information than she owns, i.e. any signal structure that depends on her own signal. The contents of the signal structure is contractible. The buyer’s willingness to pay for information depends on both the buyer’s private signal and the informational good. The seller’s revenue maximization problem is formulated as a nonlinear pricing problem, where the seller elicits buyer’s private signal using a menu of information-price pairs.\par

The nonlinear pricing problem has been extensively studied when seller is only designing a one-dimensional ``quantity'' or ``quality'' of the good (\cite{mirrlees1971exploration}, \cite{mussa1978monopoly}, \cite{maskin1984monopoly}). In contrast, this paper considers informational good, whose dimensionality is infinite. Moreover, the space is not well-ordered --- although all buyer types agree that full revelation is the best and no information is the worst, there is no consensus which of two generic signal structure is better among buyers with different private signals.\par

I fully characterize the revenue-maximizing menu in the case that: 1) seller’s private signal is binary 2) buyer’s decision problem involves a rich set of actions. I first show that the revenue-maximizing menu contains only ``simple'' signal structures. Each signal structure in the menu has only two possible realizations: one perfectly reveals one of seller’s signal, and the other partially reveals the seller’s other signal. Second, the optimal menu contains a continuum of signal structures. Since each signal structure reveals one of the seller’s signal, the menu can be divided into two classes of signal structures, each fully revealing one of the seller’s signal. All signal structures in each class are ordered by Blackwell informativeness. In equilibrium, the buyer purchases a signal structure that reveals the ex ante less likely signal (according to buyer’s private belief), namely he purchases contradictory information. The more certain the buyer is about the seller’s signal, the less information he purchases. \par

The profitability of a rich menu of signal structures can be seen by considering the buyer’s willingness to pay for signal structures. Restrict the consideration to the ``simple'' signal structures. Consider the ``Diff-in-diff'' of utility from two signal structure with different informativeness for two buyer types. Hypothetically assume that a same action is optimally chosen when observing the partially revealing signal by both buyer types with both signal structures. When this action involves few risk, i.e. the payoffs of the action is almost the same as the counterfactual optimal action following the other signal, the DID is almost zero. Otherwise when this action involves high risk, the DID is large. This DID is in fact the amount of rent that can be extracted (locally) from the type that values information more. Therefore, inducing the buyers with less sure prior (interpreted as the ``high type'' in standard nonlinear pricing framework) to purchase the signals that induces riskier action is relaxing the incentive compatibility constraint, comparing to a flat-price menu. In my setup where buyer’s decision problem is rich, such effect generically exists and a rich menu is optimal. \par

In \cref{sec:comp}, I study the comparative statics when the buyer’s private signal becomes more informative. In this case, a buyer of a given private belief about seller’s information purchases a more informative signal structure under the new optimal menu. This suggests that the monopolistic provision of information is complementary to the private information the buyer owns.

\subsection*{Related literatures}
My paper is built on the literature studying nonlinear pricing. It differs from classical works including \cite{mirrlees1971exploration}, \cite{mussa1978monopoly} and \cite{maskin1984monopoly} in the generality of available contracts for the seller. As a nature property of informational goods, I allow the monopoly seller to include any Blackwell experiments in the menu, instead of a one-dimensional quantity or quality. On the other hand, this paper also differs from the set of papers studying mechanism design with fully general space of contracts (e.g. \cite{noldeke2018implementation}). Utilizing special properties of informational good, I am able to get full characterization of revenue maximizing mechanism, as opposed to only existence results or partial characterizations in this literature. \par
My paper is closely related to the literature on selling information. It focuses on uncertainty about a payoff relevant state and heterogeneity in decision maker's belief (or equivalent buyer's private signal). \cite{horner2011selling} focus on information provider with private type (competent or not) and derived a gradual persuasion rule as the optimal disclosure of private information. \cite{bergemann2013selling} spent one section on optimal non-linear pricing mechanism for selling consumer level matching value data. \cite{esHo2007price} studied a model with seller controlling release of payoff relevant state. The seller can contract on the action of decision maker. My approach is different in the modeling of the informational good. I embedded the general definition of decision problem and information defined in \cite{blackwell1951comparison} into a monopoly pricing problem.\par
The paper closest to mine is \cite{bergemann2014selling}, which also studies optimal menu of Blackwell experiments. \cite{bergemann2014selling} provide a partial characterization of the signal structures in the optimal menu in a general setup (general state and action spaces), and a full characterization of the optimal menu with binary states and actions. My paper complements theirs in providing the full characterization with binary states and general action spaces. I show that when the action space is rich, the revenue maximizing menu contains a very rich set of different signal structures, as opposed to flat-pricing being optimal when action is binary in \cite{bergemann2014selling}.\par

There has been a large literature studying the value of information. Starting from \cite{blackwell1951comparison}, a general fair price is proved to be impossible to determine. \cite{cabrales2010entropy} shows that the Entropy function can be used to order information for a specific class of investment problem. \cite{moscarini2002law} prices information as an coefficient determining asymptotic value of repeated experiments. In my approach, I fix the decision problem but allow general prior beliefs. Also I am trying to find optimal monopoly prices instead of fair prices.\par
\vspace{0.5em}

The rest of the paper is organized as follows: In \cref{sec:set}, I set up the revenue maximization problem. In \cref{sec:thm}, I characterize the revenue maximizing menu of signal structures. In \cref{sec:proof}, I introduce the methodology for solving the optimal menu. In \cref{sec:comp}, I provide comparative statics on the distribution of buyer's beliefs. In \cref{sec:comp}, I provide numerical examples to visualize my results. I conclude in \cref{sec:con} Technical proofs omitted in the paper are provided in the Appendix.

\section{Setup}
\label{sec:set}
In this section, I set up a general model of nonlinear pricing of information.\par
\textbf{Decision problem}: The buyer of information is choosing an action $a\in A$. The payoff of each action $u(a,x)$ depends on a state $x\in X$. The buyer and seller have common prior belief of the state distribution $\pi_X\in \Delta(X)$ .\par
\textbf{Private information}: Buyer and seller observes private signal $t_B\in T_B$ and $t_S\in T_S$. $t_B$ and $t_S$ are jointly distributed according to a distribution $\pi(t_S,t_B|x)$. I assume that $t_S$ involves two realizations and is a sufficient statistic for $t_B$ with respect to $x$, namely the seller of information knows more than the buyer about the payoff relevant state. \par
	\textbf{Signal structure}: The seller can contract on the provision of a signal structure, defined as a pair $(S,g)$. $S$ is the set of signal realizations and $g(s|t_s):T_S\to \Delta(S)$ is the conditional distribution of signal. A signal structure $(S,g)$ reports seller's private signal $t_S$ with some noise. Notice that it is without loss of generality to assume that all signal structures share the same signal set $S$. As a result each signal structure can be represented by only $g$. Each contract specifies a pair $(g,p)$, the signal structure $g$ and corresponding price $p$.\par
	\textbf{Timing of the game}: 0) Nature draws a state $x$ according to $\pi_X$. 1) Before any signal is revealed, the seller specifies a menu of contracts $\left\{ g_j,p_j \right\}_{j\in J}$. ($j$ is a not necessarily countable general index) 2) The private signals $t_S$ and $t_B$ are realized according to $\pi$. 3) Buyer chooses an utility maximizing contract $j$ based on his private signal $t_B$. 4) Signal $s$ is realized according to $g_j$, and buyer picks an optimal action based on both $t_B$ and $s$.\par
	The nonlinear pricing problem is formulated as a two-step optimization problem. The first step is solving buyer's optimal choice of contract given a menu. The second step is solving the seller's revenue maximizing menu.
	\subsection{Buyer's problem}
	Given a signal structure $g_j$ and private signal $t_B$, buyer's optimal choice rule is a function $a(s):S\times T_B\to A$ that maximize:
	\begin{align*}
		\sup_{a(s,t_B)}\int u(a(s),x)g_j(s|t_S)\pi(t_S,t_B|x)\pi_X(x)\d s,t_S,x
	\end{align*}
	Define $U(g_j,t_B)$ as:
	\begin{align*}
		U(g_j|t_B)=\frac{\sup_{a(s,t_B)}\int u(a(s),x)g_j(s|t_S)\pi(t_S,t_B|x)\pi_X(x)\d s,t_S,x}{\int \pi(t_S,t_B|x)\pi_X(x)\d t_S,x}
	\end{align*}
	$U(g_j|t_B)$ is the conditional utility from observing the signal structure $g_j$. To capture the outside optimal of observing no information, let $g_{\emptyset}$ be the null signal structure and $p_{\emptyset}=0$. Then, given a menu $(g_j,p_j)$, buyer's optimal choice rule is a function $\iota(t_B): T_B\to J\bigcup \left\{ \emptyset \right\}$ that maximize:
	\begin{align*}
		\sup_{\iota(t_B)}\int \left(U(g_{\iota(t_B)},t_B)-p_{\iota(t_B)}\right)\pi(t_B|x)\pi_X(x)\d t_B,x
	\end{align*}
	\subsection{Seller's problem}
	Given buyer's optimal choice rule, the seller's optimization problem is:
	\begin{align}
		\sup_{(g_j,p_j)}&\int p_{\iota(t_B)} \pi(t_B,x)\pi_X(x)\d t_B,x 	\tag{P}	\label{eqn:P}\\
		\mathrm{s.t.}\ &
		\begin{cases}
		\iota(t_B)\in\arg\max \int \left(U(g_{\iota(t_B)},t_B)-p_{\iota(t_B)}\right)\pi(t_B|x)\pi_X(x)\d t_B,x\notag\\
		\text{$p_{\emptyset}=0$ and $g_{\emptyset}$ is null signal structure}
	\end{cases}
	\end{align}

	\subsection{A belief based representation}
	Since by assumption $t_S$ is a sufficient statistics for $t_B$ with respect to $x$, conditional on knowing $t_S$ the information contained in $t_B$ is irrelevant. Therefore $\mathrm{Prob}(x|s,t_B)=\mathrm{Prob}(x|t_S)\mathrm{Prob}(t_S|s,t_B)$. Let $\widetilde{u}(a,t_S)=\int u(a,x)\mathrm{Prob}(x|t_S)$, then
	\begin{align*}
		U(g_j,t_B)=\sup_{a(s,t_B)}\int u(a(s,t_B),x)\mathrm{Prob}(x|s,t_B)\d x=\sup_{a(s,t_B)}\int\widetilde{u}(a(s,t_B),t_S)\mathrm{Prob}(t_S|s,t_B)\d t_S
	\end{align*}
	Meanwhile, the seller's objective function in \cref{eqn:P} is $\sup_{(g_j,p_j)}\int p_{\iota(t_B)}\mathrm{Prob}(t_B)\d t_B$, and buyer's objective function is $\max_{\iota} \int (U(g_{\iota},t_B)-p_{\iota})\mathrm{Prob}(t_B)\d t_B $. These two formula don not explicitly involves state $x$.\par
	The previous analysis suggests that in the formulation of \cref{eqn:P}, in fact both buyer and seller's objective function does not explicitly involve state $x$ once we use $\widetilde{u}$ to replace $ \int u(a,x)\mathrm{Prob}(x|t_S)$. So the whole nonlinear pricing problem \cref{eqn:P} can actually be reformulated as a problem where $t_S$ is directly the payoff relevant state. Two buyer's signals $t_B$ differs in term of buyer's choice of menu only if they induce different beliefs about the distribution of $t_S$. So buyer's private signal can be equivalently summarized as his private belief about seller's signal. Here I represent an equivalent belief based representation of the problem:
	Seller's signal $t_S$ can be either $h$ or $l$. Let $\mu$ be buyer's private belief of seller's type being $h$. Let $V(\mu)=\sup_{a} E_{\mu}[\widetilde{u}(a,t_S)]$. Given each signal structure $g_j$, buyer's posterior belief observation signal realization $s$ is $\widehat{\mu}(\mu,s,g_j)=\frac{g_j(s|h)\mu}{g_j(s|h)\mu+g_j(s|l)(1-\mu)}$ by Bayes rule. Then it is obvious that $V(\widehat{\mu}(\mu,s,g_j))$ is the maximal expected utility from choosing the optimal action conditional on observing signal $s$ and $E_{s}\left[ V(\widehat{\mu}(\mu,s,g_j)) \right]=U(g_j,\mu)$. Invoke the revelation principle, the nonlinear pricing problem can be written as seller choosing direct mechanism $\mathcal{M}=\left( g_{\mu},p_{\mu} \right)$ to maximize:
\begin{align}
	\max_{\mathcal{M}}\ &\int_0^1 p_{\mu} f(\mu)d\mu      \label{eqn:seller1}\\
	s.t.\ &E_{s}\left[ V\left( \hat{\mu}(\mu,s,g_{\mu}) \right) \right]-p_{\mu}\ge E_{s}\left[ V\left( \hat{\mu}(\mu,s,g_{\mu'}) \right) \right]-p_{\mu'} \tag{IC} \\
				&E_{s}\left[ V\left( \hat{\mu}(\mu,s,g_{\mu}) \right) \right]-p_{\mu}\ge V(\mu) \tag{IR}
\end{align}
\cref{eqn:seller1} is formulated as a standard nonlinear pricing problem, except for that the seller is choosing a signal distribution $g_{\mu}$ for each buyer, as opposed to one dimensional quantity or quality in the canonical models. \par
It is worth noticing that apart from the contract space, the type space is also non-standard. Consider for example a fully revealing signal structure. The buyer with ex ante very imprecise signal has the highest willingness to pay for the signal structure. Therefore, in the belief space $[0,1]$, they types are nor linearly ordered. In this paper, I overcome the difficulties brought by the infinite dimensional contract space and non-ordered type space, by showing that 1) it is without loss of optimality to consider a one-dimensional contract space (i.e. the ``simple'' signal structures). 2) the type space can be divided into two subsets, on each of with types are linearly ordered.

\section{Characterization of the Optimal Menu}
\label{sec:thm}
\subsection{The role of rich actions}
In this section, I use a simple example to illustrate why a rich menu of signal structures can improve upon the flat-price strategy. First, consider the seller's revenue from selling the fully revealing signal structure at a flat price. The buyer's utility gain from this signal structure given private belief $\mu$ is $\mathrm{co}V(\mu)$ ($\mathrm{co}V$ is the upper concave hull of $V$ defined as in \cite{kamenica2009bayesian}). Therefore, given price $p$, all buyer types with $\mathrm{co}V(\mu)-V(\mu)\ge p$ purchases the signal. The revenue is $p\times \mathrm{Prob}(\mathrm{co}F(\mu)-V(\mu)\ge p)$. The optimal choice of price maximizes the revenue.\par

\begin{figure}[htbp]
	\centering
	\begin{minipage}{0.45\linewidth}
		\centering
	\includegraphics[width=0.9\linewidth]{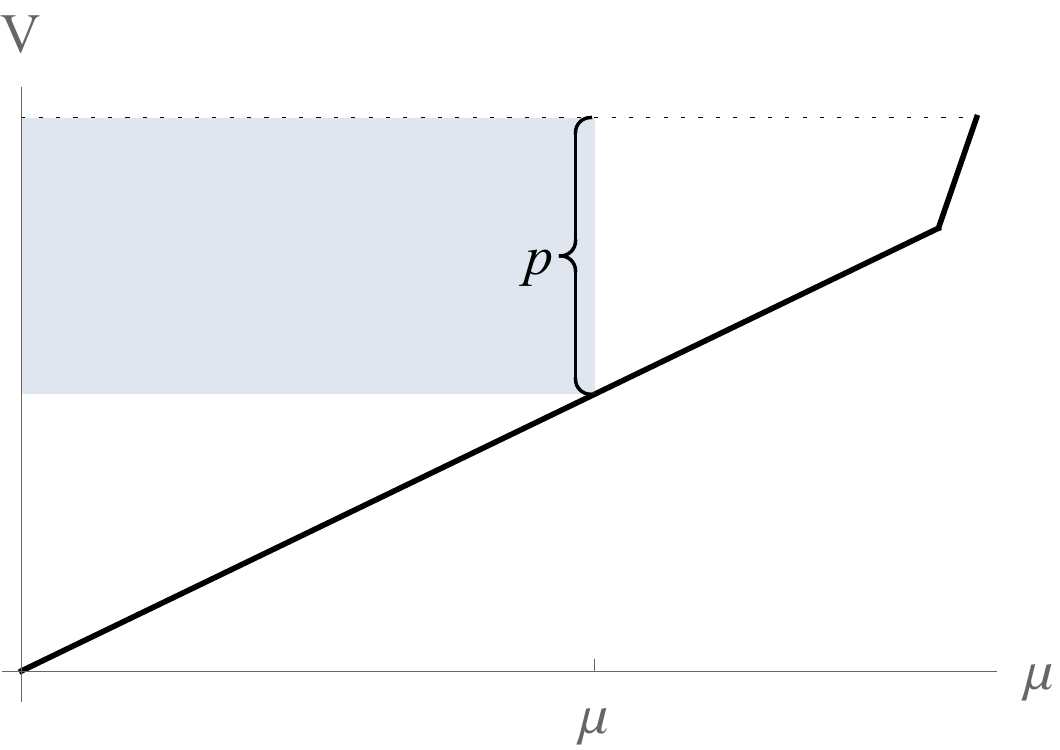}
	\caption{Optimal flat-price}
	\label{fig:ex:1}
\end{minipage}
	\begin{minipage}{0.45\linewidth}
	\centering
	\includegraphics[width=0.9\linewidth]{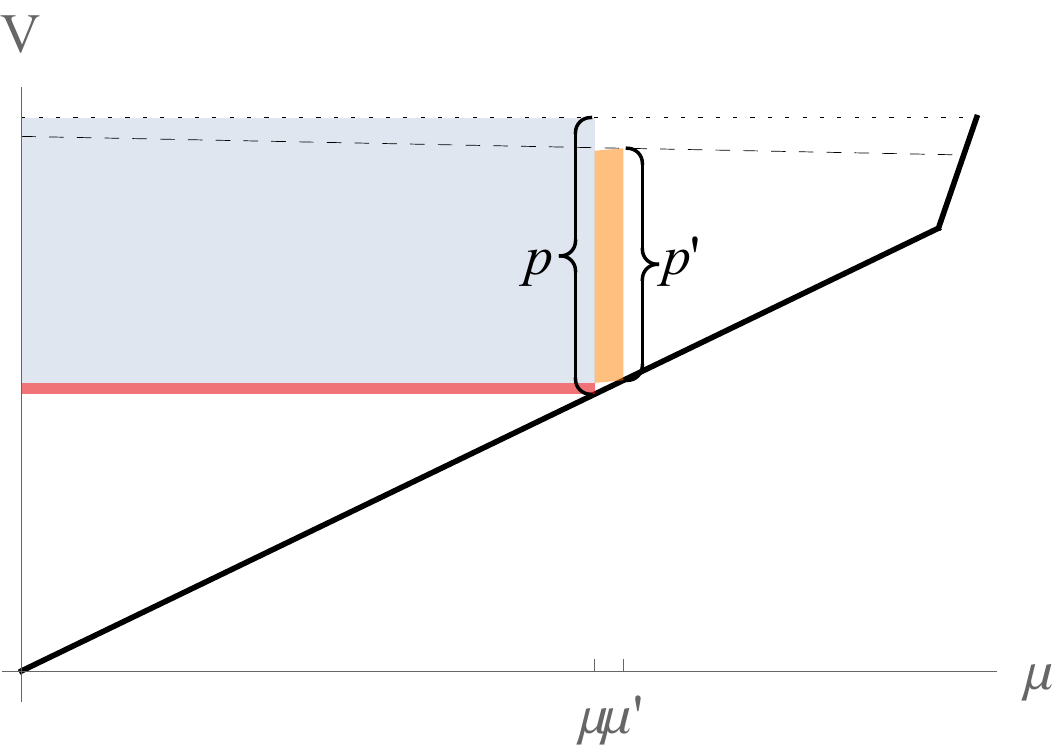}
	\caption{Discriminative pricing}
	\label{fig:ex:2}
\end{minipage}
\end{figure}

In \cref{fig:ex:1}, I show an example with a symmetric $V$ function (the black curve) which is piecewise linear (there are three kinks, which represents $4$ alternative actions). The distribution $F$ is uniform. To simplify the illustration, I only plot $\mu\in[0.5,1]$ as everything is completely symmetric around $0.5$. $p$ is the optimal flat price for the fully revealing signal structure. All buyer's with belief within $[1-\mu,\mu]$ purchases this contract. The area of blue region in \cref{fig:ex:1} represents the optimal revenue from flat-pricing.

\par
Now consider including a small set of buyer types within $[\mu,\mu']$ by introducing a second option. In this example, let the second option be a ``simple'' signal structure. It reveals state $l$ perfectly, and state $h$ imperfectly. Consider the buyer with belief $\mu'$. The posterior belief induced by the fully revealing signal is $0$ and the posterior belief $\nu$ induced by the partially revealing signal is interior. The dashed line in \cref{fig:ex:2} show the linear combination of  $(0,V(0))$ and $(\nu,V(\nu))$. Standard analysis implies that the utility of type $\mu'$ from observing this signal structure is exactly the dashed line evaluated at $\mu'$. By the individual rationality of type $\mu'$, the price of the second signal structure is $p'$.\par

The second option is more attractive than the first one to some buyer types who originally purchase the first option. For example, type $\mu$ gets zero surplus from the fully revealing signal structure, but strictly positive surplus from the second option. So to make all buyer types who originally purchase option one still purchase option one, the price of option one should be reduced to $p_1<p$ to give those buyer types more surplus. To sum up, by introducing a second option and reducing the price of the first option, the revenue gain is from including more buyer types (the area of orange region in \cref{fig:ex:2}), the revenue loss is from reducing the price for fully revealing signal structure (the area of red region in \cref{fig:ex:2}). Whether introducing the second option improves revenue depends on the comparison of the areas of the two regions.\par

Now let us calculate the revenue gain and loss from introducing the second option. A key term is the reduction in price $p$ when buyer type $\mu$ is different between the two options. Denote the fully revealing signal structure $g_f$ and the second signal structure $g_1$. The downward incentive compatibility constraint implies:
\begin{align*}
	U(g_f,\mu)-&p_1-V(\mu)\ge U(g_1,\mu)-p'-V(\mu)\\
	\implies p-p_1\ge& p-p'-\left( U(g_f,\mu)-U(g_1,\mu) \right)\\
	=&U(g_f,\mu)-V(\mu)-U(g_1,\mu')+V(\mu')-\left( U(g_f,\mu)-U(g_1,\mu) \right)\\
	=&(V(\mu')-V(\mu))-\left( U(g_1,\mu')-U(g_1,\mu) \right)\\
	\sim&  \left( V'(\mu)-\frac{\partial}{\partial \mu}U(g_1,\mu) \right)\d \mu\text{ when $\d \mu=\mu'-\mu\to 0$}
\end{align*}
let $a_{\mu}$ be the optimal choice of action when belief is $\mu$. Then the value of $V'(\mu)$ is $u(a_{\mu},r)-u(a_{\mu},l)$.
Now we calculate the value of $\frac{\partial}{\partial\mu}U(0g_1,\mu$. By construction of $g_1$, one of the signal reveals state $l$ for sure. That is to say, there exists $q$ s.t.:
\begin{align*}
	\begin{cases}
		g_1(s_0|l)=q;\ g_1(s_1|l)=1-q\\
		g_1(s_0|r)=0;\ g_1(s_1|r)=1
	\end{cases}
\end{align*}
Let $a_0$, $a_1$ be the corresponding optimal action observing signal $s_0$ and $s_1$. Then:
\begin{align*}
	U(g_1,\mu)=&\mu u(a_1,r)+(1-\mu)\left( q u(a_0,l)+(1-q)u(a_1,l) \right)\\
	\implies\frac{\partial}{\partial \mu}U(g_1,\mu)=&u(a_1,r)-\left( q u(a_0,l)+(1-q)u(a_1,l) \right)\\
	= & u(a_1,l)(1-q)
\end{align*}
The last equality is by symmetry of the problem. To sum up:
\begin{align*}
	\d p \sim \underbrace{\left( u(a_{\mu},r)-u(a_{\mu},l) \right)}_{\text{riskiness of action $a_{\mu}$}} - u(a_0,r)(1-q)
\end{align*}
The price change due to introduction of the second signal structure can be  decomposed into two terms. The first term captures the riskiness of the optimal action chosen at belief $\mu$ --- potential loss of utility when in the counterfactual state. The second negative term is proportional to $u(a_0,r)$, the utility of the wrong action chosen at state $r$. $\d p$ is smaller, namely the informational rent extracted from type $\mu$ is larger, when the first term is smaller and the second negative term is larger.\par
It is possible that $\d p$ is sufficiently small that including the second option is profitable. In fact, in the numerical example that generates \cref{fig:ex:1,fig:ex:2}, the area of the orange region is larger than the red region, namely flat-pricing is dominated by a richer menu. This analysis also illustrates the characterization of \cite{bergemann2014selling}'s binary type, binary action model. When action is binary, the value of $(u(a_{\mu},r)-u(a_{\mu},l))$ is coupled with $u(a_1,r)(1-q)$ in a particular way (in fact $a_{\mu}=a_1$). In this special case, the analysis in \cite{bergemann2014selling} proves that introducing more options is always suboptimal. However, as I have illustrated, considering a more general decision problem for buyer of information decouples the two terms determining the informational rent that can be extracted by add a second signal structure. Therefore, a rich decision problem might lead to a rich revenue maximizing menu being optimal. This intuition will be confirmed formally in \cref{thm:1}.

\subsection{Assumptions}
I need a series of technical assumptions to obtain a good form of solution.
\begin{ass}    \label{ass:mon}
    Distribution of buyer's beliefs $f(\mu)\in\mathcal{L}[0,1]$ satisfies $\forall \lambda$:
    \begin{align*}
        &\frac{f(\mu)(1-\mu)}{\lambda+F(\mu)-1}\text{ decreasing with $\mu>F^{-1}(1-\lambda)$}.\\
        &\frac{f(\mu)\mu}{\lambda+F(\mu)-1}\text{decreasing with $\mu<F^{-1}(1-\lambda)$ }.
    \end{align*}
\end{ass}

\cref{ass:mon} is an analog of the standard monotonic likelihood ratio condition. In this problem, since an uncertain belief gets most extra utility from information (a ``high type''), the type space is not linearly ordered. In stead, the willingness to pay for information as a function of belief first increases then decreases. So the type space can be divided into two intervals, on each of which the type is linearly ordered. \cref{ass:mon} modifies the standard monotonic likelihood ratio condition such that it operates on each of the two ordered type regions properly.

\begin{ass}${}$
	\label{ass:1}
	\begin{enumerate}[noitemsep]
		\item $V(\mu)\in C^{(2)}(0,1)$. There exists $\underline{\mu}<\overbar{\mu}$ and $\underline{\lambda}<\overbar{\lambda}$ in $(0,1)$ such that:
			\begin{align*}
				\forall \mu\in[\underline{\mu},\overbar{\mu}],& 
				\begin{cases}
				V(\mu)+V'(\mu)(1-\mu)+V''(\mu)\frac{\mu(1-\mu)^2}{1-\bar{\mu}} \text{ is strictly increasing}\\
				V(\mu)-V'(\mu)\mu+V''(\mu)\frac{\mu^2(1-\mu)}{\underline{\mu}} \text{ is strictly decreasing}
				\end{cases}\\
				\forall \lambda\in[\underline{\lambda},\overbar{\lambda}],&
          \begin{cases}
                \left( 1+\frac{f(\underline{\mu})\underline{\mu}}{\bar{\lambda}+F(\underline{\mu})-1} \right)  \left( V(\mu)+V'(\mu)(1-\mu) \right)+V''(\mu)\frac{\mu(1-\mu)^2}{1-\underline{\mu}}\text{ is strictly increasing}\\
                \left( 1-\frac{f(\bar{\mu})(1-\bar{\mu})}{\underline{\lambda}+F(\bar{\mu})-1} \right)\left( V(\mu)-V'(\mu)\mu \right)+V''(\mu)\frac{\mu^2(1-\mu)}{\bar{\mu}}\text{ is strictly decreasing}
            \end{cases}
			\end{align*}
		\item  $\overbar{\mu}^*$ and $\underline{\mu}^*$ are defined by $\frac{f(\bar{\mu}^*)(1-\bar{\mu}^*)}{\underline{\lambda}+F(\bar{\mu}^*)-1}=1$ and $\frac{f(\underline{\mu}^*)\underline{\mu}^*}{\bar{\lambda}+F(\underline{\mu}^*)-1}=1$. Let $\pi^*$ be the profit earned by seller using mechanism proposed in \cref{thm:1}, then:
            \begin{align*}
                \max\left\{ \bar{\mu}^*V(1)+(1-\bar{\mu}^*)V(0)-V(\bar{\mu}^*),\underline{\mu}^*V(1)+(1-\underline{\mu}^*)V(0)-V(\underline{\mu}^*) \right\}<\pi^*
            \end{align*}
	\end{enumerate}
\end{ass}

\cref{ass:1} is a sufficient condition for the optimal mechanism to contain only ``simple'' signal structures. The first part is an analog of two standard assumptions widely used in nonlinear pricing: \emph{supermodularity} condition and \emph{monotonic virtual value} condition. First of all, as I have discussed in \cref{ass:mon}, since the type space is divided into two ordered subsets, each condition need to be defined on the two sets separately in a symmetric way. The first set of monotonicity conditions states the standard supermodularity condition. The reason that I need them to be satisfied only on $\mu\in[\underline{\mu},\overbar{\mu}]$ is that when $\mu$ is outside of the interval, supermodularity is implied by the monotonic virtual value conditions. The second set of monotonicity conditions states the standard monotonic virtual value condition. The equivalence will be clear when I calculate the virtual value. $\left[ \underline{\lambda},\overbar{\lambda} \right]$ is the region for possible locations of the (endogenous) threshold where monotonicity of types switches. Finally, the second part is a non-standard condition. It implies that choosing the threshold being too extreme (outside of $[\underline{\lambda},\overbar{\lambda}]$) is dominated by the optimal mechanism proposed in \cref{thm:1}. 
\begin{remark}
    If underlying problem is symmetric ($f$ and $V$ are both symmetry around $0.5$), then $\bar{\lambda}=\underline{\lambda}=0.5$ in \cref{ass:1}, as the threshold belief where monotonicity of types switches is $0.5$. In this case, it is not hard to verify that only the supermodularity conditions are sufficient for rest two.
\end{remark}

\subsection{Optimal menu}

\begin{thm}
    The optimal mechanism solving \cref{eqn:seller1} involves signal structures each with up to two signals. There exists $\lambda\in(\underline{\lambda},\bar{\lambda})$ and $\mu^+,\mu^-$ satisfying:
    \begin{align*}
        \begin{cases}
    f(\mu^-)\mu^-+(\lambda+F(\mu^-)-1)=0\\
        f(\mu^+)(1-\mu^+)-(\lambda+F(\mu^+)-1)=0\\
        \end{cases}
    \end{align*}
    \begin{itemize}
        \item For $\mu\in[\mu^-,\mu^+]$, a signal structure fully revealing the state is sold at flat price.
        \item For $\mu\in[0,\mu^-]$, a signal structure fully revealing only state $h$ is sold. The non-conclusive signal induces posterior belief $\min\left\{ \nu,\mu \right\}$, where $\nu$ is defined by:
            \begin{align*}
                \left( 1+\frac{f(\mu)\mu}{\lambda+F(\mu)-1} \right)(V(\nu)-V(1)+V'(\nu)(1-\nu))+V''(\nu)\frac{\nu(1-\nu)^2}{1-\mu}=0
            \end{align*}
        \item For $\mu\in[\mu^+,1]$, a signal structure fully revealing only $l$ will be sold. The non-conclusive signal induces posterior belief $\max\left\{ \mu_1,\mu \right\}$, where $\mu_1 $ is defined by:
            \begin{align*}
                \left( 1-\frac{f(\mu)(1-\mu)}{\lambda+F(\mu)-1} \right)(V(\nu)-V(0)-V'(\nu)\nu)+V''(\nu)\frac{\nu^2(1-\nu)}{\mu}=0
            \end{align*}
    \end{itemize}
    \label{thm:1}
\end{thm}

\cref{thm:1} states that the revenue maximizing menu contains a rich set of simple signal structures. The signal structures in the optimal menu is simple in that each of them involves only two signal. One of the signal reveals one of the state perfectly, while the other signal partially reveals another state. For buyer types with uncertain prior belief, which values information a lot, the fully revealing signal structure is sold to them at a flat price. The remaining buyer types are ordered by the distances of their beliefs from the center region. Buyer with more uncertain prior belief purchases a more informative signal structure. For those buyer types, if the prior belief is higher on state $h$, then the buyer is assigned a signal structure fully revealing state $l$, and vice versa. This means the optimal menu assigns signal structures that precisely contradicts buyer's private belief, and imprecisely confirms buyer's private belief.\par

To illustrate the characterization in \cref{thm:1}, I calculate a numerical example:

\begin{eg}
	\label{eg:1}
	Suppose the underlying decision problem is to choose actions from $[0,1]$ to minimize a quadratic loss: $u(a,\theta)=-(x-\theta)^2$ where $\theta=\left\{ 0,1 \right\}$. It can be equality verified that $V(\mu)=\mu^2-\mu$. Let buyer's private belief about seller's private signal distribute uniformly on $[0,1]$. First, as I discussed in \cref{ass:1}, symmetry of the problem makes \cref{ass:mon,ass:1} easy to verify. The condition in \cref{thm:1} that characterizes optimal posterior induced by the non-conclusive signal in each signal structure reduces to a linear function. When $\mu>0.5$:
\begin{align*}
    1+2\mu^2-3.5\mu+(2\mu-1)\mu_1=0
\end{align*}
The revenue maximizing menu is illustrated in \ref{fig:1}:
\begin{figure}[htbp]
    \centering
    \includegraphics[width=0.6\linewidth]{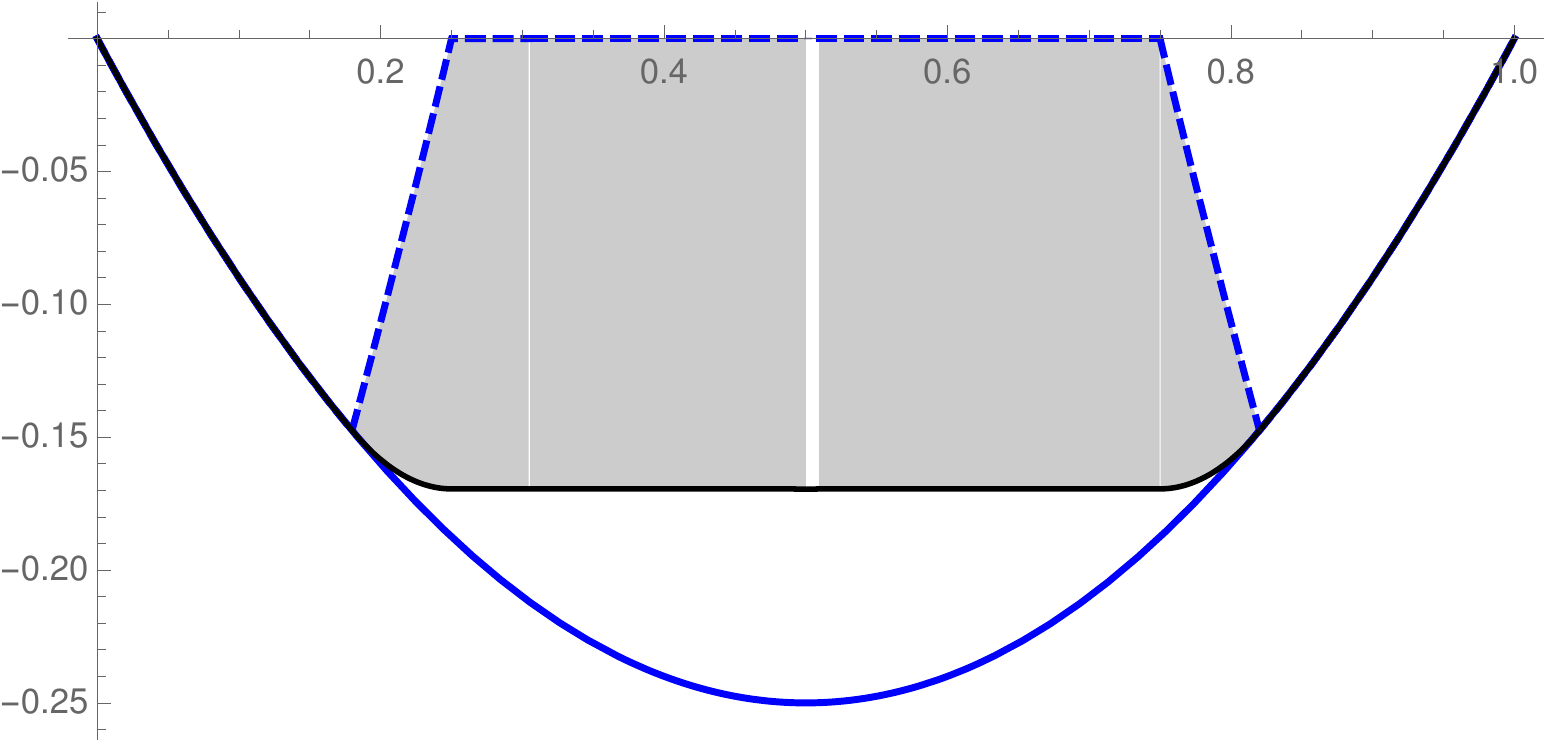}
    \caption{Optimal mechanism with quadratic utility and uniform distribution}
    \label{fig:1}
\end{figure}\par
The solid blue line is utility function on belief $V(\mu)$. The dashed line is buyer's expected payoff from the decision problem given choosing the incentive compatible signal structure-price pair from the optimal menu. The difference between dashed line and solid black line is the price charged by the seller. For buyers with belief $\mu$ roughly between $0.25$ and $0.75$, a fully revealing experiment is sold at flat price. For buyers with more extreme beliefs, a continuum of partially revealing experiments is sold. Buyers with even more extreme beliefs are excluded from the market. \par	
\end{eg}

\section{Proof methodology}
\label{sec:proof}

As is discussed, there are two main difficulties in solving \cref{eqn:seller1}. First, the type space is not linearly ordered. Second, the contract space is an infinite dimensional space. In this section, I show that main methodology I use to overcome the two difficulties and solve for the optimal menu.

\subsection{Simplification}
In this section, I discuss how the dimensionality of space of contracts can be reduced. First of all, as a standard approach in nonlinear pricing, instead of solving problem \eqref{eqn:seller1}, I study a relaxed problem:
\begin{align}
    \max_{\mathcal{M}}\ &\int_0^1p(\mu)f(\mu)d\mu     \label{eqn:seller2}\\
    s.t.\ &\frac{\partial}{\partial s_{\mu}}E_{s_{\mu}}\left[ V(\hat{\mu}(\mu,s_{\mu}))\right]\frac{d s_{\mu}}{d \mu}=p'(\mu)\tag{IC1}\\
    &p(0)=p(1)=0 \tag{IR1}
\end{align}
\cref{eqn:seller2} relaxes the global incentive compatibility constraints and individual rationality constraints in \cref{eqn:seller1} to only local ones. Since at $\mu=0$ and $\mu=1$, experiments are of no value to the decision maker, the price at these two beliefs must be zero. Therefore IR1 is a necessary condition. By a standard envelope theorem argument, global $IC$ constraint implies local $IC$ constraint (IC1). Therefore, \cref{lem:1} is quite straightforward:
\begin{lem}
    \label{lem:1}
    The solution of \cref{eqn:seller2} satisfying global $IC$ and $IR$ is a solution of \cref{eqn:seller2}.
\end{lem}
Given \cref{lem:1}, my first task is to solve the relaxed problem \eqref{eqn:seller2} and then verify the global $IC$ and $IR$. From the relaxed problem \cref{eqn:seller2}, $IC1$ can be rewritten by integrating it:
\begin{align*}
    p(\mu)=&\int_0^\mu\left( \frac{d}{d\mu}E_{s_{\mu}}\left[ V(\hat{\mu},s_{\mu}) \right]-E_{s_{\mu}}\left[\frac{\partial}{\partial \mu}V(\hat{\mu},s_{\mu}) \right]\right) d\mu\\
    =&E_{s_{\mu}}\left[ V(\hat{\mu},s_{_\mu}) \right]-V(\mu)-\int_0^{\mu} E_{s_{\mu}}\left[ \frac{\partial}{\partial \mu}(V(\hat{\mu},s_{\mu})-V(\mu)) \right]d\mu
\end{align*}
Replace $p(\mu)$ in the objective function:
\begin{align}
    \max_{\mathcal{M}}&\int_0^1 E_{s_{\mu}}\left[ V(\hat{\mu},s_{_\mu}) \right]f(\mu)d\mu-\int_0^1  E_{s_{\mu}}\left[ \frac{\partial}{\partial \mu}(V(\hat{\mu},s_{\mu})-V(\mu)) \right](1-F(\mu))d\mu     \label{eqn:seller3}\\
    s.t.\ &\int_0^1  E_{s_{\mu}}\left[ \frac{\partial}{\partial \mu}(V(\hat{\mu},s_{\mu})-V(\mu)) \right] d\mu=0 \notag
\end{align}\par
Now let's define the exact form of experiments in the menu. Let the set of possible signals be $S=\left\{ s_i \right\}$ ($i$ is general index). Each signal $s_i$ realizes in state $l$ with probability $q_i$ and in state $h$ with probability $p_i$. Thus, when signal $s_i$ is observed, posterior belief $\hat{\mu}(s_i,\mu)=\frac{p_i\mu}{p_i\mu+q_i(1-\mu)}$. The probability that signal $s_i$ realizes is $p_i\mu+q_i(1-\mu)$ from the buyer with prior $\mu$'s point of view. \cref{eqn:seller2} can be written as:

\begin{align}
    \max_{\left\{ p_i,q_i \right\}}&\int_0^1\left(\sum_i\left( p_i\mu+q_i(1-\mu) \right)V(\frac{p_i\mu}{p_i\mu+q_i(1-\mu)})-V(\mu)\right)f(\mu)d\mu    \label{eqn:seller4}\\
    &-\int_0^1\frac{\partial}{\partial \mu}\left(\sum_i\left( p_i\mu+q_i(1-\mu) \right)V(\frac{p_i\mu}{p_i\mu+q_i(1-\mu)})-V(\mu)\right)(1-F(\mu))d\mu\notag\\
    s.t.\ &\int_0^1\frac{\partial}{\partial \mu}\left(\sum_i\left( p_i\mu+q_i(1-\mu) \right)V(\frac{p_i\mu}{p_i\mu+q_i(1-\mu)})-V(\mu)\right)d\mu=0\notag\\
    &\sum_ip_i=\sum_iq_i=1\notag
\end{align}

To simplify the problem in the dimensionality of contracts in the optimal menu, I develop the following lemmas that shows that the optimal menu only includes a class of experiment with very simple form.
\begin{lem}
    The optimal menu which solves \cref{eqn:seller4} includes experiments with up to three signals, and up to one of them is partially informative.
    \label{lem:2}
\end{lem}
The intuition for proving lemma \ref{lem:2} is simple. I prove that conditional on existence of at least two interior signals in an experiment, any two interior signals should induce the same posterior belief to satisfy a first order condition. Interior signal is defined as a signal which realizes with probability within $(0,1)$ at any state. Under this condition, the first order condition involving two interior signals includes no multiplier on choice of conditional probability. Therefore, I can use the monotonicity of first order condition to show that the two interior signals must be identical. Then \cref{ass:1} guarantees that for any $\mu\in(0,1)$, this monotonicity holds. Therefore,  the optimal mechanism involves experiments with no more than three signals. And at most one of the three signals is partially informative.\par
With \cref{lem:2}, I can reduce the dimensionality of mechanism space to two (the posterior associated with the interior signal and the probability of realization of this signal). I further simplify this problem into a one-dimensional problem using the following lemma:
\begin{lem}
    The optimal mechanism which solves \cref{eqn:seller4} includes experiments with up to two signals.
    \label{lem:3}
\end{lem}
To prove \cref{lem:3}, I assume for the purpose of contradiction the existence of two fully revealing signals in an experiment. Then, I study the first order condition of an interior signal. By studying the sign of first order condition in different regions, I conclude that the first order conditions associated with the two fully revealing signals can not hold simultaneously. Therefore, I conclude that the optimal menu only includes experiments with up two signals and at most one of them is partially revealing.

\subsection{Solving the optimal mechanism}
With \cref{lem:1,lem:2,lem:3}, \cref{eqn:seller1} is simplified into a simple one-dimensional mechanism design problem with single variate. Before proceeding to writing down the reduced problem, I still want to determine what kind of experiments is sold to what kind of consumer. To be specific, I want to know the region of buyer to which experiments perfectly revealing $l$ and experiments perfectly revealing $h$ are sold.
\begin{lem}
    Let $\mu^0=F^{-1}(1-\lambda)$, 
    \begin{itemize}
        \item For $\mu\in[0,\mu^0]$, experiments revealing $h$ are sold.
        \item For $\mu\in[\mu^0,1]$, experiments revealing $l$ are sold.
    \end{itemize}
    \label{lem:4}
\end{lem}\par
With lemma \ref{lem:4}, I can write down the reduced problem with a mechanism sending signals $\left\{ H,L \right\}$ defined as following:
\begin{itemize}
    \item For $\mu\in[\mu^0,1]$, when $\theta=l$, $p(L)=q,p(H)=1-q$, when $\theta=h$, $p(L)=0,p(H)=1$.
    \item For $\mu\in[0,\mu^0]$, when $\theta=h$, $p(L)=1-p,p(H)=p$, when $\theta=l$, $p(L)=1,p(H)=0$.
\end{itemize}
I define $\Delta V(\mu,p,q)$ as the surplus of buyer type $\mu$ purchasing experiment with $p,q$ defined as before. Then the expression of $\Delta V(\mu,,q)$ is:
\begin{align*}
    \Delta V(\mu,p,q)=
    \begin{cases}
        p\mu V(1)+(\mu(1-p)+1-\mu)V\left( \frac{p\mu}{\mu(1-p)+1-\mu} \right)-V(\mu)&\text{if $\mu<\mu^0$}\\
        q(1-\mu)V(0)+(\mu+(1-q)(1-\mu))V\left( \frac{\mu}{\mu+(1-q)(1-\mu)} \right)-V(\mu)&\text{if $\mu\ge\mu^0$}
    \end{cases}
\end{align*}
Therefore the optimization problem for seller can be rewritten as:
\begin{align*}
    \max_{p(\mu),q(\mu)}\ &\int_{0}^{\mu^0}\Delta V(\mu,p(\mu),1)f(\mu)d\mu+\int_{\mu^0}^1\Delta V(\mu,1,q(\mu))f(\mu )d\mu\\
    &+\int_0^{\mu^0}(1-F(\mu))\frac{\partial}{\partial\mu}\Delta V(\mu,p(\mu),1)d\mu+\int_{\mu^0}^1(1-F(\mu))\frac{\partial}{\partial\mu}\Delta V(\mu,1,q(\mu))d\mu\\
    s.t.\ & \int_0^{\mu^0}\frac{\partial}{\partial\mu}\Delta V(\mu,p(\mu),1)d\mu+\int_{\mu^0}^1\frac{\partial}{\partial\mu}\Delta V(\mu,1,q(\mu))d\mu=0
\end{align*}
Let $\lambda_1$ be Lagrangian multiplier on the constraint, then the optimality condition requires $\lambda_1$ exactly being $\lambda$ defined in \cref{thm:1}. The FOCs are:
\begin{alignat}{2}
    &\left( 1-\frac{(1-\mu)f(\mu)}{\lambda+F(\mu)-1} \right)\left( V(\mu_1)-V(0)-\mu_1V'(\mu_1) \right)+V''(\mu_1)\frac{\mu_1^2(1-\mu_1)}{\mu}=0
    \label{eqn:FOC1}\\
    &\left( 1+\frac{\mu f(\mu)}{\lambda+F(\mu)-1} \right)(V(\mu_2)-V(1)+(1-\mu_2)V'(\mu_2))+V''(\mu_2)\frac{\mu_2(1-\mu_2)^2}{1-\mu}=0
    \label{eqn:FOC2}
\end{alignat}
$\mu_1$ is the induced belief when the other signal reveals $l$, $\mu_2$ is the induced belief when the other signal reveals $h$.Then to establish uniqueness result, let's study the two FOCs more carefully. First define $\mu^+$ and $\mu^-$:
    \begin{align*}
        \begin{cases}
    f(\mu^-)\mu^-+(\lambda+F(\mu^-)-1)=0\\
        f(\mu^+)(1-\mu^+)-(\lambda+F(\mu^+)-1)=0\\
        \end{cases}
    \end{align*}
$[0,1]$ can be devided into four regions:\par
\textbf{Region 1:} $\mu\in(0,\mu^-)$\\
In this region, by \cref{lem:4} $h$ is revealed. Therefore first order condition for $p$ is characterized by \cref{eqn:FOC2}. Since $\mu<\mu^-$, $1+\frac{\mu f(\mu)}{\lambda+F(\mu)-1}>0$. Therefore \cref{eqn:FOC2} might have solution. Moreover, since $\mu$ can be set arbitrarily close to $\mu^-$, there exist a positive mass of $\mu$ such that \cref{eqn:FOC2} has solution.\par
\textbf{Region 2:} $\mu\in[\mu^-,\mu^0]$\\
In this region, I still use \cref{eqn:FOC2}. However, in this region $1+\frac{\mu f(\mu)}{\lambda+F(\mu)-1}\le0$. Therefore \cref{eqn:FOC2} has no solution with $\mu_2>0$. The only possibility that this $FOC$ might hold is when $p=1$ and induces a positive multiplier.  So in this region, the fully revealing experiment is sold.\par
\textbf{Retion 3}: $\mu\in[\mu^0,\mu^+]$\\
In this region, by \cref{lem:4} $l$ is revealed. Therefore first order condition for $q$ is characterized by \cref{eqn:FOC1}. Since $\mu\le \mu^+$, $1-\frac{(1-\mu)f(\mu)}{\lambda+F(\mu)-1}\le0$. Therefore \cref{eqn:FOC1} has no solution with $\mu_1<1$. The only possibility that $FOC$ might hold is when $q=1$ and induces a positive multiplier. So in this region, the fully revealing experiment is sold.\par
\textbf{Region 4}: $\mu\in(\mu^+,1)$\\
In this region, by \cref{lem:4} $l$ is revealed. Therefore $FOC$ for $q$ is characterized by \cref{eqn:FOC1}. Since $\mu>\mu^+$, $1-\frac{\mu f(\mu)}{\lambda+F(\mu)-1}>0$, \cref{eqn:FOC1} might have solution. Similar to present argument, there exist a positive mass of $\mu$ such that \cref{eqn:FOC1} has solution.\par
 It's necessary to verify the single crossing difference condition and monotonicity condition to make local ICs sufficient for global ICs and IRs. Cross derivatives of $\Delta V$ are:
\begin{alignat}{2}
    &\frac{\partial^2}{\partial\mu\partial p}\Delta V=-V(\mu_2)+V(1)-V'(\mu_2)(1-\mu_2)-V''(\mu_2)\frac{\mu_2(1-\mu_2)^2}{1-\mu}\label{eqn:SCD1}\\
    &\frac{\partial^2}{\partial\mu \partial q}\Delta V=V(\mu_1)-V(0)-V'(\mu_1)\mu_1+V''(\mu_1)\frac{\mu_1^2(1-\mu_1)}{\mu}\label{eqn:SCD2}
\end{alignat}
\cref{eqn:SCD1} applies when $\mu<\mu^0$. According to \cref{ass:1}, $\mu<\bar{\mu}$, therefore \cref{eqn:SCD1} is monotonically decreasing. Meanwhile at $\mu_2=1$, this term is $0$. Therefore this term is positive for any $\mu_2<1$. That is to say, single crossing difference condition implies an increasing $p(\mu)$ when $\mu<\mu^0$. \cref{eqn:SCD2} applies when $\mu>\mu^0$. According to \cref{ass:1}, $\mu>\underline{\mu}$. Therefore \cref{eqn:SCD2} is monotonically increasing.  At $\mu_1=0$, this term is $0$. Therefore this term should be negative at any $\mu_1>0$. This is to say, single crossing difference condition implies an decreasing $q(\mu)$ when $\mu>\mu^0$.

\section{Comparative Statics}
\label{sec:comp}
In this section, I do a comparative statics analysis by shifting the distribution of buyer's private types. From this point on, I make the following symmetry assumption:
\begin{ass}
    The environment is symmetric i.e.$V(\mu)=V(1-\mu)$, $f(\mu)=f(1-\mu)$.
    \label{ass:sym}
\end{ass}
Given assumption \ref{ass:sym}, the environment is symmetric. Therefore there always exists a symmetric optimal menu. A symmetric menu implies $\lambda=0.5$. Therefore, the unknown parameter $\lambda$ is uniquely determined and I can define an order on distribution.
\begin{defi}
    $\mathcal{F}=\left\{ f|f\in\Delta[0,1], f(x)=f(1-x) \text{, satisfying assumption \ref{ass:mon}} \right\}$, $\forall f,g\in\mathcal{F}$, $F,G$ are corresponding CDF. We define $f$ being more dispersed than $g$ if for $\mu<\mu^-_g$:
    \begin{align*}
        \frac{f(\mu)}{0.5-F(\mu)}\ge\frac{g(\mu)}{0.5-G(\mu)}
    \end{align*}
    $\mu^-_g$ defined by solution of $g(\mu)+G(\mu)-0.5=0$.
    \label{defi:1}
\end{defi}\par
Holding the decision problem fixed, when distribution of buyers become more dispersed, of course the buyers with more extreme beliefs become relatively more important. Thus seller has incentive to cut price and include more buyers to buy some products. Let's have a formal proof for this intuition. Since the problem is now totally symmetric, I can only focus on $\mu\in(0,0.5)$. By theorem \ref{thm:1}, the interval of buyers to whom the fully revealing experiment is sold is determined only by distribution. Given distribution $g$, the end point $\mu^-$ is solution to:
\begin{align*}
    1=\frac{g(\mu)\mu}{0.5-G(\mu)}
\end{align*}
Therefore, by definition of dispersive order, the equality might not hold at $\mu^-$ if $f$ is more dispersed than $g$. Also by monotonicity assumption on distribution, corresponding $\mu^-$ must decreases.\par
The interval of buyers to whom at least a partially informative experiment is sold is jointly determined by distribution and decision problem. Given distribution $g$, the end point is solution to:
\begin{align*}
    (1-\frac{g(\mu)\mu}{0.5-G(\mu)})(V(\mu)-V(1)+V'(\mu)(1-\mu))+V''(\mu)\mu(1-\mu)=0
\end{align*}
By definition of dispersive order, the equality might not hold at original point if $f$ is more dispersed than $g$. Assume that the end point is different from original one, I want to argue that it must be smaller. If this is not the case, then $\mu_1(\mu)$ as a function under distribution $f$ must cross $\mu_1(\mu)$ under $g$ at least once. However this is not possible according to first order condition. Because at a same $\mu$, changing distribution to a more dispersed one only changes the coefficient before a strict negative term. Noticing that here I didn't prove continuity of mechanism. So a more strict treatment is needed in a formal proof. To summarize, I have the following proposition.
\begin{prop}
    Given underlying decision problem $V$, when distribution of buyer's belief become more dispersed, the interval to which fully revealing experiment is sold is expanding. $\mu^-$ decreases and $\mu^+$ increases. The interval to which at least partially revealing experiment is sold is expanding.
    \label{prop:2}
\end{prop}\par
Combining \cref{prop:2} and the no crossing argument, I can easily derive the following proposition:
\begin{prop}
    Given underlying decision problem $V$, when distribution of buyer's belief become more dispersed, any buyer holding a prior belief $\mu$ will be sold a Blackwell more informative experiment in optimal mechanism.
    \label{prop:3}
\end{prop}
\cref{prop:3} states that when distribution of buyer's belief become more dispersed, all buyer type are sold a better experiment which generates higher payoff from the decision problem. I am also interested whether this higher payoff generates higher surplus for buyers. Let's start by rewriting surplus for a specific buyer $\Delta V(E(\mu),\mu)-p(\mu)$. Local incentive compatibility constraint implies:
\begin{align*}
    &P'(\mu)=\frac{\partial}{\partial E}\Delta V(E,\mu)\frac{d}{d\mu}E(\mu)\\
    \Rightarrow\ &\frac{d}{d\mu}\left(\Delta V(E(\mu),\mu)-P(\mu)\right)=\frac{\partial}{\partial \mu}\Delta V(E,\mu)\\
    \Rightarrow\ &\Delta V(E,\mu)-P(\mu)=\int_0^{\mu}\Delta V_{\mu}(E,\nu)d\nu
\end{align*}
Let's again focus only on $\mu\in(0,0.5)$ by symmetry. Now let $p(\mu)$ be the probability defining experiments for optimal mechanism under distribution $g$. Let's $p^1(\mu)$ be the corresponding mechanism under distribution $f$ and $f$ is more dispersed than $g$. Therefore by \cref{prop:3} $p(\mu)\le p^1(\mu)$. Then:
\begin{align*}
    \Delta V(p(\mu),\mu)-P(\mu)&=\int_0^{\mu}\Delta V_{\mu}(p(\mu),\nu)d\nu\\
    &=\int_0^{\mu}\int_0^{p(\nu)}\frac{\partial^2}{\partial\mu\partial p} \Delta V(p,\nu)dp d\nu\\
    &\le  \int_0^{\mu}\int_0^{p^1(\nu)}\frac{\partial^2}{\partial\mu\partial p} \Delta V(p,\nu)dp d\nu\\
    &= \Delta V(p^1(\mu),\mu)-P(\mu)
\end{align*}
The inequality is implied by single crossing different condition: $\frac{\partial^2}{\partial\mu\partial p}\Delta V(p,\mu)>0$. To sum up, I have the following proposition.
\begin{prop}
    Given underlying decision problem $V$, when distribution of buyer's belief become more dispersed, any buyer holding a belief $\mu$ will be weakly better off in optimal mechanism.
    \label{prop:4}
\end{prop}

\begin{eg}
	\label{eg:2}
	Starting from the same setup in \cref{eg:1}. Fix the same belief and modify distribution by rotating pdf of the uniform distribution around the critical points defined as in \cref{defi:1}. It's easy to verify that this rotation operation satisfies the dispersive order. Then the resulting optimal mechanisms are depicted in \cref{fig:2}:
\begin{figure}[h]
    \centering
    \includegraphics[width=0.6\linewidth]{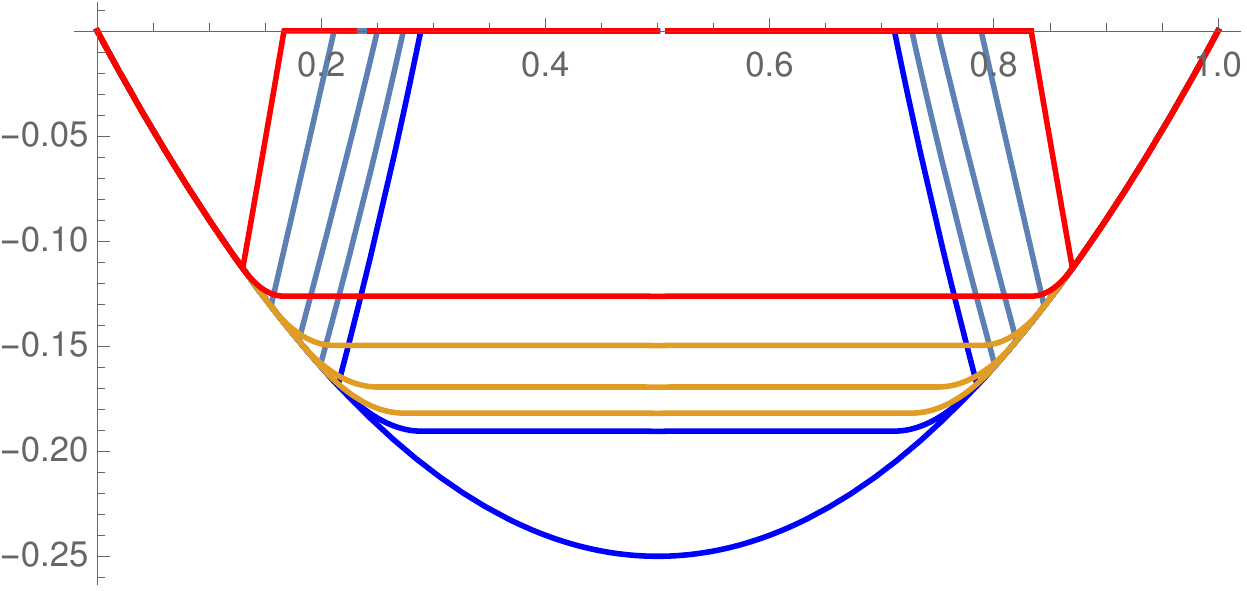}
    \caption{Comparative statics}
    \label{fig:2}
\end{figure}\par
I calculate $5$ different levels of dispersiveness in this example. When the buyer's private belief distribution is most dispersed, the optimal menu is characterized by the red curve. When the buyer's private belief distribution is most condensed, the optimal menu is characterized by the blue curve. It is clear that for any given buyer's private belief, the more dispersed the belief distribution is, the better signal structure he purchases from the revenue maximizing menu (the upper red curve is higher than the upper blue curve). Meanwhile, he gets more surplus from trading ( the lower red curve is higher than the lower blue curve).
\end{eg}

\section{Conclusion}
\label{sec:con}
In this paper, I study the optimal nonlinear pricing of information in the environment of selling information to a buyer whole is making a decision and observed private signal about the payoffs. I focus on the case that the buyer's decision problem involves a rich set of different actions, and the seller's information is binary. I show that under some regularity conditions, the seller's revenue maximizing menu contains a rich set of ``simple'' signal structures. The ``simple'' signal structures each contains only two signals, one of which perfect revealing a state. When the buyer's private belief is very uncertain, he purchases a fully revealing signal structure at a flat price. When his private belief is more certain, he purchases a signal structure that perfectly contradicts his prior belief. The more certain he is, the less informative signal structure he purchases.\par

I also provide a comparative statics result on the dispersiveness of belief distribution. When the belief distribution becomes more dispersed, targeting least informed seller types become less profitable so the seller include a wider interval of buyer types into the menu and also provide the fully revealing experiment to a wider interval of buyer types. All buyer types are offered a Blackwell more informative experiment. Corresponding price decreases and buyer with a certain belief enjoys a higher surplus.\par

\newpage
\appendix
\section{Proofs for \cref{sec:proof}}
\subsection{Proof of Lemma \ref{lem:2}}
Let's assign $\lambda$ as the Lagrangian multiplier on the integral constraint. Then let's replace some signal $j$ using $\sum_ip_i=\sum_iq_i=1$. Thus the whole problem will be:
\begin{align*}
    L=&\int_0^1\left(\sum_i\left( p_i\mu+q_i(1-\mu) \right)V(\frac{p_i\mu}{p_i\mu+q_i(1-\mu)})-V(\mu)\right)f(\mu)d\mu\\
    &+\int_0^1\frac{\partial}{\partial \mu}\left(\sum_i\left( p_i\mu+q_i(1-\mu) \right)V(\frac{p_i\mu}{p_i\mu+q_i(1-\mu)})-V(\mu)\right)(1-F(\mu)-\lambda)d\mu\\
\end{align*}
We take FOC for $p_i$:
\begin{align*}
    &(f(\mu)\mu+(\lambda+F(\mu)-1))\left( V(\mu_i)+V'(\mu_i)(1-\mu_i) \right)+(\lambda+F(\mu)-1)V''(\mu_i)\frac{\mu_i(1-\mu_i)^2}{1-\mu}\\
    =&(f(\mu)\mu+(\lambda+F(\mu)-1))\left( V(\mu_j)+V'(\mu_j)(1-\mu_j) \right)+(\lambda+F(\mu)-1)V''(\mu_j)\frac{\mu_j(1-\mu_j)^2}{1-\mu}+\gamma_p^+-\gamma_p^-
\end{align*}
Here $\gamma_p^+$ is the multiplier for $p_j\le0$ and $\gamma_p^-$ is the multiplier for $p_i\le0$. Similarly let's take FOC for $q_i$, with multipliers defined in same way:
\begin{align*}
     &(f(\mu)(1-\mu)-(\lambda+F(\mu)-1))\left( V(\mu_i)-V'(\mu_i)\mu_i \right)-(\lambda+F(\mu)-1)V''(\mu_i)\frac{\mu_i^2(1-\mu_i)}{\mu}\\
    =&(f(\mu)(1-\mu)-(\lambda+F(\mu)-1))\left( V(\mu_j)+V'(\mu_j)\mu_j \right)-(\lambda+F(\mu)-1)V''(\mu_j)\frac{\mu_j^2(1-\mu_j)}{\mu}+\gamma_q^+-\gamma_q^-   
\end{align*}
Now let's assume $\mu_j$ is an interior belief, i.e. $p_j,q_j\in(0,1)$. Thus if $\mu_i$ is also an interior belief, for the mechanism to be optimal, I require FOCs to be held with $\gamma_p^+=\gamma_p^-=\gamma_q^+=\gamma_q^-=0$. Thus, the resulting $\mu_i,\mu_j$ is determined by function:
\begin{align*}
    &H_1(\mu_i)= (f(\mu)\mu+(\lambda-F(\mu)-1))\left( V(\mu_i)+V'(\mu_i)(1-\mu_i) \right)+(\lambda+F(\mu)-1)V''(\mu_i)\frac{\mu_i(1-\mu_i)^2}{1-\mu}\\
    &H_2(\mu_i)= (f(\mu)(1-\mu)-(\lambda-F(\mu)-1))\left( V(\mu_i)-V'(\mu_i)\mu_i \right)-(\lambda+F(\mu)-1)V''(\mu_i)\frac{\mu_i^2(1-\mu_i)}{\mu}
\end{align*}
If both functions are monotonic on (0,1), then all interior $\mu_i$s must have the same value. To investigate this, I divide $[0,1]$ into four regions and discuss $\mu$ case by case. Also I use assumption \ref{ass:1} to restrict the value of $\lambda$.\par
\textbf{Region 1:$\mu\in[\underline{\mu},F^{-1}(1-\lambda)]$}\\
When $\mu\in\left[ \underline{\mu},F^{-1}(1-\lambda) \right]$:
\begin{align*}
    f(\mu)(1-\mu)-(\lambda+F(\mu)-1)>-(\lambda+F(\mu)-1)\ge0
\end{align*}
 By convexity of $V$: 
\begin{align*}
    V(\mu_i)-V'(\mu_i)\mu_i\text{ weakly decreasing}
\end{align*}
Then according to part one of assumption \ref{ass:1}:
\begin{align*}
    &V(\mu_i)-V'(\mu_i)\mu_i+V''(\mu_i)\frac{\mu_i^2(1-\mu_i)}{\mu}\\
    =&\left(V(\mu_i)-V'(\mu_i)\mu_i+V''(\mu_i)\frac{\mu_i^2(1-\mu_i)}{\underline{\mu}} \right)\frac{\underline{\mu}}{\mu}+\left( 1-\frac{\underline{\mu}}{\mu} \right)(V(\mu_i)-V'(\mu_i)\mu_i)
\end{align*}
strictly decreasing. Because both terms are decreasing and the coefficients are positive. Thus:
\begin{align*}
    H_1(\mu_i)=&f(\mu)(1-\mu)\underbrace{(V(\mu_i)-V'(\mu_i)\mu_i)}_{\text{decreasing}}\\
    &+\underbrace{-(\lambda+F(\mu)-1)}_{\text{positive}}\underbrace{\left(V(\mu_i)-V'(\mu_i)\mu_i+V''(\mu_i)\frac{\mu_i^2(1-\mu_i)}{\mu}\right)}_{\text{decreasing}}
\end{align*}
is strictly decreasing. Therefore, given $\mu_j$ being interior, when $\mu\in[\underline{\mu},F^{-1}(1-\lambda)]$, interior $\mu_i$ must equal to $\mu_j$.\par
\textbf{Region 2:$\mu\in[F^{-1}(1-\lambda),\bar{\mu}]$}\\
When $\mu\in[F^{-1}(1-\lambda),\bar{\mu}]$:
\begin{align*}
    f(\mu)\mu+(\lambda+F(\mu)-1)>\lambda+F(\mu)-1
\end{align*}
By convexity of $V$:
\begin{align*}
    V(\mu_i)+V'(\mu_i)(1-\mu_i)\text{ weakly increasing}
\end{align*}
Then according to part one of assumption \ref{ass:1}:
\begin{align*}
    &V(\mu_i)+V'(\mu_i)(1-\mu_i)+V''(\mu_i)\frac{\mu_i(1-\mu_i)^2}{1-\mu}\\
    =&\left( V(\mu_i)+V'(\mu_i)(1-\mu_i)+V''(\mu_i)\frac{\mu_i(1-\mu_i)^2}{1-\bar{\mu}} \right)\frac{1-\bar{\mu}}{1-\mu}+\frac{\bar{\mu}-\mu}{1-\mu}(V(\mu_i)+V'(\mu_i)(1-\mu_i))
\end{align*}
strictly increasing. From the same argument as before. Therefore:
\begin{align*}
    H_2(\mu_i)=&f(\mu)(1-\mu)\underbrace{(V(\mu_i)+V'(\mu_i)(1-\mu_i))}_{\text{increasing}}\\
    &+\underbrace{(\lambda+F(\mu)-1)}_{\text{positive}}\underbrace{\left( V(\mu_i)+V'(\mu_i)(1-\mu_i)+V''(\mu_i)\frac{\mu_i(1-\mu_i)^2}{1-\mu} \right)}_{\text{increasing}}
\end{align*}
is strictly increasing. Therefore given $\mu_j$ being interior, when $\mu\in\left[ F^{-1}(1-\lambda),\bar{\mu} \right]$, interior $\mu_i$ must equal to $\mu_j$.\par
Then let's study $\mu$ close to $0$ or $1$. According to part three of assumption \ref{ass:1}, by optimality $\lambda\in(\underline{\lambda},\bar{\lambda})$.\par
\textbf{Region 3:$\mu\in(0,\underline{\mu}]$}\\
Since $\lambda>\underline{\lambda}$, by part two of assumption \ref{ass:1}:
\begin{align*}
    &\left( 1+\frac{f(\mu)\mu}{\lambda+F(\mu)-1} \right)(V(\mu_i)+V'(\mu_i)(1-\mu_i))+V''(\mu_i)\frac{\mu_i(1-\mu_i)^2}{1-\mu}\\
    =& \left(\left( 1+\frac{f(\mu)\mu}{\lambda+F(\mu)-1} \right)(V(\mu_i)+V'(\mu_i)(1-\mu_i))+V''(\mu_i)\frac{\mu_i(1-\mu_i)^2}{1-\underline{\mu}}\right)\frac{1-\underline{\mu}}{1-\mu}\\
    +&\frac{\underline{\mu}-\mu}{1-\mu} \left( 1+\frac{f(\mu)\mu}{\lambda+F(\mu)-1} \right)(V(\mu_i)+V'(\mu_i)(1-\mu_i))
\end{align*}
We know that the second term is weakly increasing (positive coefficient). Let's deal with the first term:
\begin{align*}
    &\left( 1+\frac{f(\mu)\mu}{\lambda+F(\mu)-1} \right)(V(\mu_i)+V'(\mu_i)(1-\mu_i))+V''(\mu_i)\frac{\mu_i(1-\mu_i)^2}{1-\underline{\mu}}\\
    =&\left( 1+\frac{f(\underline{\mu})\underline{\mu}}{\bar{\lambda}+F(\underline{\mu})-1} \right)(V(\mu_i)+V'(\mu_i)(1-\mu_i))+V''(\mu_i)\frac{\mu_i(1-\mu_i)^2}{1-\underline{\mu}}\\
    &+\left( \frac{f(\mu)\mu}{\lambda+F(\mu)-1}-\frac{f(\underline{\mu})\underline{\mu}}{\bar{\lambda}+F(\underline{\mu})-1} \right)(V(\mu_i)+V'(\mu_i)(1-\mu_i))
\end{align*}
The first term is increasing according to assumption \ref{ass:1}. The second term is increasing because the coefficient is positive given $\lambda<\bar{\lambda}$ and $\mu\le\underline{\mu}$. Noticing the term I investigated is $H_1(\mu_i)$ times some constant coefficient. Therefore given $\mu_j$ being interior, when $\mu\in(0,\underline{\mu}]$, interior $\mu_i$ must equal to $\mu_j$.\par
\textbf{Region 4:$\mu\in[\bar{\mu},1)$}\\
    Since $\lambda<\bar{\lambda}$, by part two of assumption \ref{ass:1},  I have:
    \begin{align*}
        &\left( 1-\frac{f(\mu)(1-\mu)}{\lambda+F(\mu)-1} \right)(V(\mu_i)-V'(\mu_i)\mu_i)+V''(\mu_i)\frac{\mu_i^2(1-\mu_i)}{\mu}\\
        =&\left( \left( 1-\frac{f(\mu)(1-\mu)}{\lambda+F(\mu)-1} \right)(V(\mu_i)-V'(\mu_i)\mu_i)+V''(\mu_i)\frac{\mu_i^2(1-\mu_i)}{\bar{\mu}}  \right)\frac{\bar{\mu}}{\mu}\\
        &+\frac{\mu-\bar{\mu}}{\mu}  \left( 1-\frac{f(\mu)(1-\mu)}{\lambda+F(\mu)-1} \right)(V(\mu_i)-V'(\mu_i)\mu_i) 
    \end{align*}
    We know that the second term is weakly decreasing (positive coefficient). Let's deal with the first term:
    \begin{align*}
        &\left( 1-\frac{f(\mu)(1-\mu)}{\lambda+F(\mu)-1} \right)(V(\mu_i)-V'(\mu_i)\mu_i)+V''(\mu_i)\frac{\mu_i^2(1-\mu_i)}{\bar{\mu}}\\
        =& \left( 1-\frac{f(\bar{\mu})(1-\bar{\mu})}{\underline{\lambda}+F(\bar{\mu})-1} \right)(V(\mu_i)-V'(\mu_i)\mu_i)+V''(\mu_i)\frac{\mu_i^2(1-\mu_i)}{\mu}\\
        &-\left( \frac{f(\mu)(1-\mu)}{\lambda+F(\mu)-1}-\frac{f(\bar{\mu})(1-\bar{\mu})}{\underline{\lambda}+F(\bar{\mu})-1} \right)(V(\mu_i)-V'(\mu_i)\mu_i)
    \end{align*}
    The first term is strictly decreasing according to assumption \ref{ass:1}. The second term is decreasing because the coefficient is positive given $\lambda>\underline{\lambda}$ and $\mu\ge\bar{\mu}$. Noticing the term I investigated is $H_2(\mu_i)$ times some constant coefficient. Therefore I can conclude that given $\mu_j$ being interior, when $\mu\in[\bar{\mu},1)$, interior $\mu_i$ must equal to $\mu_j$.\par
To sum up, I've proved that signals inducing interior beliefs must induce the same belief. Thus the only possibility is boundary beliefs i.e. the case where $p_i,q_i=0,1$. First, it's straight forward from the existence of another interior signal, $p_i,q_i=1$ is impossible. Because that will drive all other signals to inducing extreme beliefs. There I only need to discuss case where $p_i,q_i=0$.\par
When $p_i=0$, $\mu_i=0$. When $q_i=0$, $\mu_i=1$. Therefore, assuming there exists an signal inducing interior belief, the possible experiment satisfying optimality must only induce one interior belief and up to two extreme beliefs. Assuming there doesn't exist an signal inducing interior belief, then the only possibility is that the experiment induces two extreme beliefs.\par
To sum up, optimal mechanism solving problem \eqref{eqn:seller4} includes experiments with up to one partially informative signal. We proved lemma \ref{lem:2}.
\subsection{Proof of Lemma \ref{lem:3}}
Let's still utilize the FOCs derived in the proof of lemma \ref{lem:2}. They can be written as:
\begin{align*}
    &H_1(\mu_i)=H_1(\mu_j)+\gamma_p^+-\gamma_p^-\\
    &H_2(\mu_i)=H_2(\mu_j)+\gamma_q^+-\gamma_q^-
\end{align*}
I want to show that one of the three signals will never appear in optimal mechanism. To investigate this, I divide $[0,1]$ into four regions and discuss $\mu$ case by case. Given a $\lambda$, let's first define $\mu^-,\mu^0,\mu^+$ as:
\begin{align*}
    \begin{cases}
        f(\mu^-)\mu^-+(\lambda+F(\mu^-)-1)=0\\
        f(\mu^+)(1-\mu^+)-(\lambda+F(\mu^+)-1)=0\\
        \lambda+F(\mu^0)-1=0
    \end{cases}
\end{align*}
By assumption part three of \ref{ass:1}, $\underline{\mu}<\mu^-<\mu^+<\bar{\mu}$.\par
\textbf{Region 1:$\mu\in(0,\mu^-)$}\\
We assume that there exists an interior signal inducing $\mu_j$
\begin{itemize}
    \item If $\mu_i=1$, then FOC implies:
        \begin{align*}
            &\underbrace{(f(\mu)\mu+\lambda+F(\mu)-1)}_{\text{negative}}V(1)=H_1(\mu_j)
        \end{align*}
    \item If $\mu_i=0$, then FOC implies:
        \begin{align*}
            \underbrace{(f(\mu)\mu+\lambda+F(\mu)-1)}_{\text{negative}}(V(0)+V'(0))=H_1(\mu_j)-\gamma_p^-
        \end{align*}
    \item If both of them are true, then:
        \begin{align*}
            \underbrace{(f(\mu)\mu+\lambda+F(\mu)-1)}_{\text{negative}}\underbrace{(V(0)+V'(0)-V(1))}_{\text{negative}}=-\gamma_p^-
        \end{align*}
        This is impossible. This is saying that when $\mu\in(0,\mu^-)$, there will be up to two signals in optimal mechanism.
\end{itemize}\par
\textbf{Region 2:$\mu\in(\mu^+,1)$}\\
We assume that there exists and interior signal inducing $\mu_j$:
\begin{itemize}
    \item If $\mu_i=1$, then FOC implies:
        \begin{align*}
            (f(\mu)(1-\mu)-(\lambda+F(\mu)-1)(V(1)-V'(1))=H_2(\mu_j)-\gamma_q^-
        \end{align*}
    \item If $\mu_i=0$, then FOC implies:
        \begin{align*}
            (F(\mu)(1-\mu)-(\lambda+F(\mu)-1))V(0)=H_2(\mu_j)
        \end{align*}
    \item If both of them are true, then:
        \begin{align*}
            \underbrace{(F(\mu)(1-\mu)-(\mu+F(\mu)-1))}_{\text{negative}}\underbrace{V(1)-V'(1)-V(0)}_{\text{negative}}=-\gamma_q^-
        \end{align*}
        This is impossible. This is saying that when $\mu\in(\mu^+,1)$ there will be up to two signals in optimal mechanism.
\end{itemize}\par
\textbf{Region 3:$\mu\in[\mu^-,\mu^0]$}\\
In this region:
\begin{align*}
    f(\mu)(1-\mu)-(\lambda+F(\mu)-1)>\ge(\lambda+F(\mu)-1)\ge0
\end{align*}
Consider:
\begin{align*}
    G_2(\mu_i)=V(\mu_i)-V'(\mu_i)\mu_i-V(0)+V''(\mu_i)\frac{\mu_i(1-\mu_i)^2}{1-\mu}
\end{align*}
By assumption \ref{ass:1},$G_2$ is decreasing. It's also easy to see that $G_2(0)=0$. Thus $G_2(\mu_i)<0$ for $\mu_i>0$.\par
We assume that there exists two fully informative signals inducing $\mu_1=0$, $\mu_2=1$ and now let's consider the FOC of choosing an interior signal $\mu_3$. By $\mu_1=0,\mu_2=1$, $p_1>0,q_1=0$, $p_2=0,q_2>0$. Thus, increasing $q_i$ when using $\mu_i$ as reference will not trigger shadow costs. Choose $\mu_1$ as reference, by FOC:
\begin{align*}
    H_2(\mu_i)=H_2(0)
\end{align*}
However this is not possible for $\mu_3>0$. Because:
\begin{align*}
    H_2(\mu_i)-H_2(0)=-(\lambda+F(\mu)-1)G_2(\mu_i)+f(\mu)(1-\mu)(V(\mu_i)-V'(\mu_i)\mu_i-V(0))<0
\end{align*}
Thus when there are two fully informative signals, the optimal mechanism only includes perfect experiment for $\mu\in[\mu^-,\mu^0]$.\par
\textbf{Region 4:$\mu\in[\mu^0,\mu^+]$}\\
In the region:
\begin{align*}
    f(\mu)\mu+(\lambda+F(\mu)-1)\ge\lambda+F(\mu)-1\ge0
\end{align*}
Consider:
\begin{align*}
    G_1(\mu_i)=H_1(\mu_i)-H_1(1)=V(\mu_i)+V'(\mu_i)(1-\mu_i)-V(1)+V''(\mu_i)\frac{\mu_i^2(1-\mu_i)}{\mu}
\end{align*}
By assumption \ref{ass:1}, $G_1$ is increasing. It's also easy to see that $G_1(1)=0$. Thus $G_1(\mu_i)<0$ for $\mu_i<1$.\par
We assume that there exists two fully informative signals. Now let's consider the FOC for choosing an interior signal inducing $\mu_i$. Similar to the previous argument, increasing $p_i$ when using $\mu_2$ as reference will not trigger a shadow cost. 
\begin{align*}
    H_1(\mu_i)=H_1(1)
\end{align*}
However this is not possible for $\mu_i<1$. Because:
\begin{align*}
    H_1(\mu_i)-H_1(q)=(\lambda+F(\mu)-1)G_1(\mu_i)+F(\mu)\mu(V(\mu_i)+V'(\mu_i)(1-\mu_i)-V(1))<0
\end{align*}
Thus when there are two fully informative signals, the optimal mechanism only includes perfect experiment for $\mu\in[\mu^0,\mu^+]$.\par
\subsection{Proof of lemma \ref{lem:4}}
The only point at which experiment switches type is a point assigned with a fully revealing experiment. Then by monotonicity of mechanism, there are only two possible patterns. Either experiments revealing $0$ are sold to low prior buyers, or are sold to high prior buyers. \par
However, in the proof of lemma \ref{lem:3}, when $\mu\in[\mu^-,\mu^0]$, optimal experiment with an interior signal can not reveal $l$. When $\mu\in[\mu^0,\mu^+]$, optimal experiment with interior signal can not reveal $h$. Thus if experiments assigned to these two region are not fully revealing, then experiments revealing $l$ will be assigned to buyers with higher $\mu$ and vice versa.\par
Now the only undetermined case is when $\mu\in[\mu^-,\mu^+]$, only fully revealing experiments are sold. Let's look at the single crossing difference condition on $q_i$ when the other signal reveals $l$:
\begin{align*}
    &\frac{\partial^2}{\partial\mu \partial q_i}\left( (\mu+q_i(1-\mu))v(\frac{\mu}{\mu+q_i(1-\mu)})+(1-q_i)(1-\mu)V(0)-V(\mu) \right) \\
    =&V(\mu_i)-V(0)-V'(\mu_i)\mu_i+V''(\mu_i)\frac{\mu_i^2(1-\mu_i)}{\mu}
\end{align*}
From assumption \ref{ass:1}, for $\mu>\underline{\mu}$, the SCD is negative. Thus, it can not be the case that experiments fully revealing $l$ is sold to buyers with $\mu<\mu^0$ and increase to a fully revealing experiment.\par
A same argument can be made for for the SCD of $p_i$ and we can conclude that lemma \ref{lem:4} is true.
\newpage
\bibliographystyle{apalike2}
\bibliography{report}

\begin{thebibliography}{}

\bibitem[Bergemann \& Bonatti, 2013]{bergemann2013selling}
Bergemann, D. \& Bonatti, A. (2013).
\newblock Selling cookies.

\bibitem[Bergemann et~al., 2014]{bergemann2014selling}
Bergemann, D., Bonatti, A., \& Smolin, A. (2014).
\newblock Selling experiments: Menu pricing of information.

\bibitem[Blackwell et~al., 1951]{blackwell1951comparison}
Blackwell, D. et~al. (1951).
\newblock Comparison of experiments.
\newblock In {\em Proceedings of the second Berkeley symposium on mathematical
  statistics and probability}, volume~1  (pp.\ 93--102).

\bibitem[Cabrales et~al., 2010]{cabrales2010entropy}
Cabrales, A., Gossner, O., \& Serrano, R. (2010).
\newblock Entropy and the value of information for investors.
\newblock {\em Available at SSRN 1720963}.

\bibitem[Eso \& Szentes, 2007]{esHo2007price}
Eso, P. \& Szentes, B. (2007).
\newblock The price of advice.
\newblock {\em The Rand Journal of Economics}, 38(4), 863--880.

\bibitem[Horner \& Skrzypacz, 2011]{horner2011selling}
Horner, J. \& Skrzypacz, A. (2011).
\newblock Selling information.

\bibitem[Kamenica \& Gentzkow, 2009]{kamenica2009bayesian}
Kamenica, E. \& Gentzkow, M. (2009).
\newblock {\em Bayesian persuasion}.
\newblock Technical report, National Bureau of Economic Research.

\bibitem[Maskin \& Riley, 1984]{maskin1984monopoly}
Maskin, E. \& Riley, J. (1984).
\newblock Monopoly with incomplete information.
\newblock {\em The RAND Journal of Economics}, 15(2), 171--196.

\bibitem[Mirrlees, 1971]{mirrlees1971exploration}
Mirrlees, J.~A. (1971).
\newblock An exploration in the theory of optimum income taxation.
\newblock {\em The review of economic studies}, 38(2), 175--208.

\bibitem[Moscarini \& Smith, 2002]{moscarini2002law}
Moscarini, G. \& Smith, L. (2002).
\newblock The law of large demand for information.
\newblock {\em Econometrica}, 70(6), 2351--2366.

\bibitem[Mussa \& Rosen, 1978]{mussa1978monopoly}
Mussa, M. \& Rosen, S. (1978).
\newblock Monopoly and product quality.
\newblock {\em Journal of Economic theory}, 18(2), 301--317.

\bibitem[N{\"o}ldeke \& Samuelson, 2018]{noldeke2018implementation}
N{\"o}ldeke, G. \& Samuelson, L. (2018).
\newblock The implementation duality.
\newblock {\em Econometrica}, 86(4), 1283--1324.

\end{thebibliography}
\end{document}